\documentclass[preprint,aps]{revtex4-1}
\pdfoutput=1

\usepackage{amsmath,amssymb,amsfonts,dcolumn,color,graphicx,graphics,latexsym,placeins,epsfig}
\usepackage{epsfig}
\usepackage{bm}
\usepackage{slashed}
\usepackage{latexsym}
\usepackage{natbib}
\usepackage{url}
\usepackage{dcolumn}
\usepackage{color}
\usepackage{amsfonts,amssymb,amsmath}
\usepackage{graphicx,epsfig}
\usepackage{psfrag}
\usepackage{subfigure}
\usepackage{tabularx}
\usepackage{hyperref}
\hypersetup{colorlinks=true}

\newcommand{\be}{\begin{equation}}
\newcommand{\ee}{\end{equation}}
\newcommand{\ba}{\begin{eqnarray}}
\newcommand{\ea}{\end{eqnarray}}

\begin{document}

\title{Vector gauge boson radiation from compact binary systems in a gauged $L_\mu-L_\tau$ scenario}
\author{Tanmay Kumar Poddar$^{1}$\footnote{tanmay@prl.res.in}}
\author{Subhendra Mohanty$^{1}$\footnote{mohanty@prl.res.in}}
\author{Soumya Jana$^{1,2}$\footnote{Soumya.Jana@etu.unige.ch}}
\affiliation{${}^{1}$ {\it Theoretical Physics Division, Physical Research Laboratory, Ahmedabad 380009, India} }

\affiliation{${}^{2}$ {\it D\'epartement de Physique Th\'eorique, Universit\'e de Gen\`eve, 24 quai Ernest Ansermet, 1211Gen\`eve 4, Switzerland} }


\begin{abstract}
The orbital period of a compact binary system decays mainly due to quadrupole gravitational radiation, which agrees with the observation to within one percent. Other types of radiation such as ultralight scalar or pseudoscalar radiation, massive vector boson radiation also contribute to the decay of orbital period as long as the mass of the emitted particle is less than the orbital frequency of the compact binary system. We obtain an expression of the energy loss due to the radiation of massive vector field from the neutron star-neutron star and neutron star-white dwarf binaries. Due to large chemical potential of the degenerate electrons, neutron stars have large muon charge.  We derive the energy loss due to $U(1)_{L_\mu-L_\tau}$ gauge boson radiation from the binaries. For the radiation of vector boson, the mass is restricted by $M_{Z^\prime}<\Omega \simeq10^{-19}eV$ are the orbital frequencies of the compact star binaries. Using the formula of orbital period decay, we obtain constraints on the coupling constant of the gauge boson in the gauged $L_\mu-L_\tau$ theory for the four compact binary systems. For vector gauge boson muon coupling we find that for $M_{Z^\prime}<10^{-19}eV$, constraints on the coupling constant is $g<\mathcal{O}(10^{-20})$. We also obtain the exclusion plots of the massive vector proca field and the gauge field which can couple to muons. 
\end{abstract}


\maketitle

\section{Introduction} 
Decrease in the orbital period with time of the Hulse-Taylor (HT) binary pulsar (PSR B1913+16) provided the first indirect evidence of gravitational wave radiation \cite{hulse,taylor,weisberg}. Although the decay of the orbital period is due to mainly the quadrupole gravitational radiation \cite{peters}, radiation of other massless or ultralight scalar or pseudoscalar particles \cite{mohanty, hook, mairi, tanmay} can contribute about one percent of the observed decay of the orbital period \citep{newre}. Some other recent studies also explain such excess in orbital period decay \cite{mahapatra}. In this paper we calculate the orbital energy loss due to radiation of proca vector boson and massive vector gauge boson of $L_\mu-L_\tau$ anomaly free gauge theory \cite{foot,he,r,heeck} from the four compact binary (neutron star-neutron star, neutron star-white dwarf) systems.

The standard model (SM) of particle physics is a $SU(3)_c\times SU(2)_L\times U(1)_Y$ gauge theory and it remains invariant under four global symmetries corresponding to the lepton numbers of the three lepton families and the baryon number. These are not the gauge symmetries but one can construct three combinations in an anomaly free way and they can be gauged in the standard model. These gauge symmetries are $L_e-L_\mu$, $L_e-L_\tau$ and $L_\mu-L_\tau$. The $L_e-L_\tau$ and $L_e-L_\mu$ long range forces from the electrons can be probed in neutrino oscillation experiments \cite{masso,anjan,amol,agarwalla}. The $L_\mu-L_\tau$ gauge force is not generated in a macroscopic body like the earth and the Sun and it can not be probed in the neutrino oscillation experiments. In this paper, we point out that neutron star (NS) can have large charge of muons and, therefore, the neutron star-neutron star (NS-NS) binaries and neutron star-white dwarf (NS-WD) binaries can radiate ultralight $L_\mu-L_\tau$ vector gauge bosons.

Besides neutrons there are electrons, protons and muons in lower fraction inside a neutron star. There are around $10^{55}$ number of muons compared to about $10^{57}$ number of neutrons \cite{dutt,potekhin,goriely,chamel,pearson,new} in a typical old neutron star. The main uncertainties in our following calculations are from the chemical potential and muon content in NS, which should be at most a factor of two \cite{new,garani,bell}. 

For massive vector gauge boson radiation from the NS-NS, and NS-WD binaries, the orbital frequency of the binary orbit should be greater than the mass of the particle which restricts the mass spectrum of the massive gauge boson to $M_{Z^\prime}<10^{-19}eV$. A $L_\mu-L_\tau$ gauge boson ($Z^\prime$) exchange between muons of the neutron star gives rise to the Yukawa type potential $V(r)=\frac{g^2}{4\pi r}e^{-M_{Z^\prime} r}$. The range $\lambda$ of the force is determined by $\lambda=\frac{1}{M_{Z^\prime}}$. For emission of this ultra light vector gauge boson of mass $M_{Z^\prime}<10^{-19}eV$ from NS-NS and NS-WD binaries, the lower bound of the range of this force is $\lambda=1/M_{Z^\prime}>10^{12} m$. This ultra light mass or nearly massless gauge boson can mediate long range fifth force between the neutron stars of the binary system. Since there is no muon charge for white dwarf, the fifth force for NS-WD binaries are zero.
In this paper we show that ultra light vector $L_\mu-L_\tau$ gauge bosons can be radiated from the NS-NS and NS-WD binaries which contribute to the decay in orbital period.

The paper is organized as follows. In section II, we estimate the number of muons inside a neutron star. In section III, we derive an expression for the energy loss due to proca vector field radiation. In section IV, we derive the energy loss due to massive vector gauge boson radiation. In section V, we obtain constraints on the gauge couplings in $L_\mu-L_\tau$ gauge, for vector gauge boson radiation from two NS-NS binaries (PSR B1913+16: Hulse Taylor binary pulsar \cite{hulse,taylor,weisberg} and PSR J0737-3039: double pulsars \cite{kramer}) and two NS-WD binaries (PSR J0348+0432 \cite{john}, and PSR J1738+0333 \cite{paulo}). We also obtain the exclusion plots of vector boson muon coupling for the proca and the gauge field from the four compact binaries. In section VI, we summarize and discuss our results.

In this paper we have used the natural system of units: $\hbar=c=1$, and $G=1/M^2_{pl}$.
\section{Estimation of muon content inside a neutron star}
The chemical potential of relativistic degenerate electrons in NS is
\begin{equation}
\mu_e=(m^2_e+k^2_{fe})^\frac{1}{2}=\Big[m^2_e+(3\pi^2\rho Y_e)^\frac{2}{3}\Big]^\frac{1}{2},
\end{equation}
where $m_e$ is the mass of the electron, $k_f$ is the Fermi momentum, $\rho$ is the nucleon number density and $Y_	e$ is the electron fraction. From the charge neutrality of the neutron star, $Y_p=Y_e+Y_\mu$ and $Y_n+Y_p=1$. Above the nuclear matter density, when $\mu_e$ exceeds the mass of muon $(\sim 105MeV$, non-relativistic), electrons can convert into muons at the edge of the Fermi sphere. So $e^-\rightarrow \mu^-+\nu_e +\bar{\nu_\mu}$, $p+\mu^-\rightarrow n+\nu_\mu$, and $n\rightarrow p+\mu^-+\bar{\nu_\mu}$ may be energetically favourable. Hence, both muons and electrons can stay in neutron star and stabilize through beta equilibrium. Thus the $\beta$ stability condition becomes
\begin{equation}
\mu_n-\mu_p=\mu_e=\mu_\mu=\Big[m^2_\mu+(3\pi^2\rho Y_\mu)^\frac{2}{3}\Big]^\frac{1}{2},
\end{equation}
where $Y_\mu$ is the muon fraction inside the neutron star \cite{feng}. Muon decay $(\mu^-\rightarrow e^-+\bar{\nu_e}+\nu_\mu)$ inside the neutron star is prohibited by Fermi statistics. The Fermi energy of the electron is roughly $100MeV$ (relativistic) whereas the Fermi energy of the muon is roughly $30MeV$ (non relativistic). Hence the muon decay cannot take place as the energy levels of the electron are all filled up to the Fermi surface and the final state electron is Fermi blocked. For white dwarf, the Fermi energy of muon is very small ($\sim 1eV$) and Fermi suppression does not really apply. Thus muon decay is not obstructed in white dwarfs.

From the beta equilibrium condition the chemical potentials of muons and electrons inside the neutron star are equal which implies,
\begin{equation}
\rho_\mu=\frac{m^3_e}{3\pi^2}\Big[1+\frac{(3\pi^2\rho Y_e)^\frac{2}{3}}{m^2_e}-\frac{m^2_\mu}{m^2_e}\Big]^\frac{3}{2}.
\label{eq:den}
\end{equation}
The electron fraction $(Y_e)$ is given as \cite{pearson}
\begin{equation}
Y_e=\frac{p_1+p_2\rho+p_6\rho^{3/2}+p_3\rho^{p_7}}{1+p_4\rho^{3/2}+p_5\rho^{p_7}},
\end{equation}
where $p$'s are the parameters which can take different values for different QCD equation of states. Assuming there are $10^{57}$ number of nucleons, the nucleon number density is $\rho=0.238 fm^{-3}$ and $Y_e=0.052$ (here we put the values of $p$ parameters for BSK24 \cite{pearson} equation of state). From Eq.~(\ref{eq:den}) we obtain the muon number density $\rho_\mu=3.11\times 10^4MeV^3$. Hence the total number of muons inside the neutron star is $\rho_\mu\times \frac{4}{3}\pi R^3=1.67\times 10^{55}$ where we assume the radius of the neutron star is $R=10Km$. In the following, we take the muon number as $N=10^{55}$.

\section{Energy loss due to radiation of massive proca vector field coupling with muons}

If there is a mismatch between the observed period loss of the binary system and its theoretical  prediction from the gravitational quadrupole radiation, then other particles may also be radiated from the binaries which gives a hint of new physics. Neutron stars have large number of muon charges ($N\approx 10^{55}$) and $Z^\prime$ massive proca vector boson can be emitted from the NS in addition to the gravitational radiation, contributing to the observed orbital period decay. A NS of typical size $10Km$ can be treated as a point source, because the Compton wavelength of radiation $(\lambda=10^{12}m)$ is much larger than the size of NS. We will treat the radiation of massive $Z^\prime$ vector bosons from the NS classically. The classical current of muons $J^\mu$ in the NS is determined from the Kepler orbits and assuming the interaction vertex as $gZ^\prime_\mu J^\mu$, where $g$ is the coupling constant. Therefore, the rate of massive $Z^\prime$ boson radiation is given by
\begin{equation}
d\Gamma =g^2\sum^3_{\lambda=1}[J^\mu(k^\prime)J^{\nu*}(k^\prime)\epsilon^\lambda_\mu(k)\epsilon^{\lambda*}_\nu(k)]2\pi\delta(\omega-\omega^\prime)\frac{d^3 k}{(2\pi)^3 2\omega},
\end{equation}
where $J^\mu(k^\prime)$ is the Fourier transform of $J^\mu(x)$ and $\epsilon^\lambda_\mu(k)$ is the polarization vector of massive vector boson. The polarization sum is given as
\begin{equation}
\sum^3_{\lambda=1}\epsilon^\lambda_\mu(k)\epsilon^{\lambda*}_\nu(k)=-g_{\mu\nu}+\frac{k_\mu k_\nu}{M^2_{Z^\prime}}.
\label{eq:pol_sum}
\end{equation}
Therefore, the emission rate is
\begin{equation}
\begin{split}
 d\Gamma = & \frac{g^2}{2(2\pi)^2}\int\Big[-|J^\mu(\omega^\prime)|^2+\frac{1}{M^2_{Z^\prime}}\Big(|J^0(\omega^\prime)|^2 \omega^2+J^i(\omega^\prime)J^{j*}(\omega^\prime)k_i k_j+2J^0(\omega^\prime) J^{i*}(\omega^\prime)k_0k_i\Big)\Big]\\ 
 & \times\delta(\omega-\omega^\prime)\omega\Big(1-\frac{M^2_{Z^\prime}}{\omega^2}\Big)^{\frac{1}{2}}d\omega d\Omega_k.
\end{split}
\end{equation}
The momentum four vector of the $Z^\prime$ boson is $k_\mu=(\omega,-\vec{k})$, $k_i=|\vec{k}|\hat{n_i}$ and $k_j=|\vec{k}|\hat{n_j}$. The third term in the first bracket will not contribute anything because
\begin{equation}
\int \hat{n_i}d\Omega_k=0, \hspace{1cm}\int \hat{n_i}\hat{n_j}d\Omega_k=\frac{4\pi}{3}\delta_{ij}.
\end{equation}
Therefore, the rate of energy loss due to massive $Z^\prime$ boson radiation is
\begin{equation}
\begin{split}
\frac{dE}{dt}=&\frac{g^2}{2\pi}\int\Big[-|J^0(\omega^\prime)|^2+|J^i(\omega^\prime)|^2+\frac{\omega^2}{M^2_{Z^\prime}}|J^0(\omega^\prime)|^2+\frac{\omega^2}{3M^2_{Z^\prime}}|J^i(\omega^\prime)|^2\Big(1-\frac{M^2_{Z^\prime}}{\omega^2}\Big)\Big]\\
&\times\delta(\omega-\omega^\prime)\omega^2\Big(1-\frac{M^2_{Z^\prime}}{\omega^2}\Big)^{\frac{1}{2}}d\omega.
\end{split}
\label{eq:first}
\end{equation}
The current density for the binary stars is written as
\begin{equation}
J^\mu(x)=\sum_{b=1,2}Q_b \delta^3(\textbf{x}-\textbf{x}_b(t))u^\mu_b,
\label{eq:current}
\end{equation}
where $b=1,2$ denotes labelling of the two stars in the binary system. $Q_b$ is the total charge of the NS due to muons and $\textbf{x}_b(t)$ denotes the location of the NS. $u^\mu_b=(1,\dot{x}_b,\dot{y}_b,0)$ is the non relativistic four velocity in the x-y plane of the Kepler's orbit. A Kepler orbit in the x-y plane can be written in the parametric form as
\begin{equation}
x=a(\cos\xi-e), \hspace{2cm}y=a\sqrt{1-e^2}\sin\xi, \hspace{2cm}\Omega t=\xi-e\sin\xi,
\label{eq:parametric}
\end{equation}
where $e$ is the eccentricity, $a$ is the semi major axis of the elliptic orbit, and $\Omega=G[\frac{m_1+m_2}{a^3}]^\frac{1}{2}$ is the fundamental frequency. The angular velocity is not constant in an eccentric orbit, which means that the Fourier expansion must sum over the harmonics $n\Omega$ of the fundamental. The Fourier transform of Eq.~(\ref{eq:current}) for the spatial part of $J^\mu(\omega^\prime)$ with $\omega^\prime=n\Omega$ is
\begin{equation}
J^i(\omega^\prime)=\int  \frac{1}{T} \int^T_0 dt e^{in\Omega t}\dot{x}^i_b(t)\sum_{b=1,2} Q_b d^3 \textbf{x}^\prime e^{-i\textbf{k}^\prime. \textbf{x}^\prime} \delta^3(\textbf{x}^\prime-\textbf{x}_b(t)).
\label{eq:Ji_omega}
\end{equation}
We expand $e^{i\textbf{k}^\prime.\textbf{x}^\prime}=1+i\textbf{k}^\prime.\textbf{x}^\prime+...$ and retain the leading order term as $\textbf{k}^\prime.\textbf{x}^\prime\sim \Omega a\ll 1$ for binary star orbits. Hence, Eq.~(\ref{eq:Ji_omega}) becomes
\begin{equation}
J^i(\omega^\prime)=\frac{Q_1}{T}\int^T_0 dt e^{in\Omega t}\dot{x}^i_1(t)+\frac{Q_2}{T}\int^T_0 dt e^{in\Omega t}\dot{x}^i_2(t).
\end{equation}
In the centre of mass (c.o.m) coordinates we have $\textbf{x}^i_1=\frac{m_2\textbf{x}^i}{m_1+m_2}=\frac{M}{m_1}\textbf{x}^i$ and $\textbf{x}^i_2=-\frac{m_1\textbf{x}^i}{m_1+m_2}=-\frac{M}{m_2}\textbf{x}^i$. $M=m_1m_2/(m_1+m_2)$ is the reduced mass of the compact binary system.
Hence, we rewrite the spatial part of the current density as
\begin{equation}
J^i(\omega^\prime)=\frac{1}{T}\Big(\frac{Q_1}{m_1}-\frac{Q_2}{m_2}\Big)M\int^T_0 dt e^{in\Omega t}\dot{x^i}(t).
\end{equation}
The Fourier transform of the velocity in the Kepler orbit can be evaluated as follows.
\begin{eqnarray}
\dot{x_n}&=&\frac{1}{T}\int^T_0 e^{i\Omega nt}\dot{x}dt \nonumber\\
&=& \frac{\Omega}{2\pi}\int^{2\pi}_0 e^{in(\xi-e\sin\xi)}(-a\sin\xi)d\xi,
\label{eq:xdot_omega}
\end{eqnarray}
where $T=2\pi/\Omega$ and, from Eq.~(\ref{eq:parametric}), we have used the fact that $\dot{x}dt=-a\sin\xi d\xi$. Similarly, we write 
\begin{equation}
\dot{y_n}=\frac{1}{T}\int^T_0 e^{i\Omega nt}\dot{y}dt.
\end{equation}
From Eq.~(\ref{eq:parametric}) we use the fact that $\dot{y}dt=a\sqrt{1-e^2}\cos\xi d\xi$ and we obtain
\begin{eqnarray}
\dot{y_n}&=&\frac{\Omega a\sqrt{1-e^2}}{2\pi}\int^{2\pi}_0 e^{in(\xi-e\sin\xi)}\cos\xi d\xi \nonumber\\
&=& \frac{\Omega a\sqrt{1-e^2}}{2\pi e}\int^{2\pi}_0 e^{in(\xi-e\sin\xi)} d\xi.
\label{eq:ydot_omega}
\end{eqnarray}
Using the identity of the Bessel function
\begin{equation}
J_n(z)=\frac{1}{2\pi}\int^{2\pi}_0 e^{i(n\xi-z\sin\xi)}d \xi,
\label{eq:J_n_Z}
\end{equation}
in Eqs.~(\ref{eq:xdot_omega}) and (\ref{eq:ydot_omega}), we obtain the velocities in Fourier space as
\begin{equation}
\dot{x}_n=-ia\Omega J^\prime_n(ne), \hspace{2cm} \dot{y}_n=\frac{a\sqrt{1-e^2}\Omega}{e} J_n(ne),
\end{equation}
where the prime over the Bessel function denotes derivative with respect to the argument.
Hence, we have
\begin{eqnarray}
J^x(\omega^\prime)&=&\Omega \Big(\frac{Q_1}{m_1}-\frac{Q_2}{m_2}\Big)M \frac{1}{2\pi}\int^T_0 dt e^{in\Omega t}\dot{x}^i(t) \nonumber\\
&=&-ia\Omega\Big(\frac{Q_1}{m_1}-\frac{Q_2}{m_2}\Big)M J^\prime_n(ne).
\end{eqnarray}
Similarly,
\begin{equation}
J^y(\omega^\prime)=\Omega\Big(\frac{Q_1}{m_1}-\frac{Q_2}{m_2}\Big)M \frac{a\sqrt{1-e^2}}{e}J_n(ne).
\end{equation}
Hence, the square of the spatial part of $J^\mu(\omega^\prime)$ becomes
\begin{eqnarray}
|J^i(\omega^\prime)|^2&=&|J^x(\omega^\prime)|^2+|J^y(\omega^\prime)|^2\nonumber\\
&=&a^2\Omega^2 M^2\Big(\frac{Q_1}{m_1}-\frac{Q_2}{m_2}\Big)^2\Big[J^{\prime^2}_n(ne)+\frac{(1-e^2)}{e^2}J^2_n(ne)\Big].
\label{eq:ji_omega}
\end{eqnarray}
From Eq.~(\ref{eq:current}), we have the temporal component of $J^\mu(\omega^\prime)$ as
\begin{equation}
J^0(\omega)=\frac{1}{2\pi}\int e^{i\textbf{k}^\prime.\textbf{x}^\prime} e^{-i\omega t}\sum_{b=1,2} Q_b  \delta^3(\textbf{x}^\prime-\textbf{x}_b(t))d^3 \textbf{x}^\prime dt.
\end{equation}
Going to the c.o.m frame, the integral results in
\begin{equation}
J^0(\omega)= (Q_1+Q_2)\delta (\omega)+ i M \left(\frac{Q_1}{m_1} - \frac{Q_2}{m_2}\right)(k_x x(\omega) + k_y y(\omega))+ \mathcal{O}((\textbf{k}.\textbf{r})^2),
\label{eq:jzero}
\end{equation}
where $x(\omega)=aJ'_n(ne)/n$ and $y(\omega)=i a \sqrt{1-e^2} J_n(ne)/{ne}$ are the Fourier transforms of the orbital coordinates. The first term in Eq.~(\ref{eq:jzero}) does not contribute due to the delta function $\delta(\omega)$. Therefore considering the second term as the leading order contribution, we obtain 
\begin{equation}
\vert J^0(\omega) \vert^2= \frac{1}{3} a^2 M^2 \Omega^2 \left(1- \frac{M_{Z^\prime}^2}{n^2\Omega^2}\right) \left(\frac{Q_1}{m_1}-\frac{Q_2}{m_2}\right)^2\left(J'^2_n(ne)+ \frac{1-e^2}{e^2} J^2_n(ne) \right),
\label{eq:j0_omega}
\end{equation}
where we have used $<k_x^2>=<k_y^2>=k^2/3$ and $\omega= n \Omega$. Using Eqs.~(\ref{eq:ji_omega}) and (\ref{eq:j0_omega}) in Eq.~(\ref{eq:first}), we obtain the rate of energy loss
\begin{eqnarray}
\frac{dE}{dt}&=&\frac{g^2}{3\pi}a^2M^2\Big(\frac{Q_1}{m_1}-\frac{Q_2}{m_2}\Big)^2\left[\frac{\Omega^6}{M^2_{Z^\prime}}\sum_{n>n_0}n^4\Big[J^{\prime^2}_n(ne)+\frac{(1-e^2)}{e^2}J^2_n(ne)\Big]\Big(1-\frac{n^2_0}{n^2}\Big)^\frac{3}{2}\right. \nonumber\\
&&\left.+
\Omega^4\sum_{n>n_0}n^2\Big[J^{\prime^2}_n(ne)+\frac{(1-e^2)}{e^2}J^2_n(ne)\Big]\Big(1-\frac{n^2_0}{n^2}\Big)^\frac{1}{2}\Big(1+\frac{1}{2}\frac{n^2_0}{n^2}\Big)\right].
\label{eq:vecto}
\end{eqnarray}
Where $n_0=M_{Z^\prime}/\Omega<1$.

We define
\begin{eqnarray}
K_1(n_0,e)=\sum_{n>n_0}n^4\Big[J^{\prime^2}_n(ne)+\frac{(1-e^2)}{e^2}J^2_n(ne)\Big]\Big(1-\frac{n^2_0}{n^2}\Big)^\frac{3}{2},
\label{eq:k1}
\end{eqnarray}
and
\begin{eqnarray}
K_2(n_0,e)=\sum_{n>n_0}n^2\Big[J^{\prime^2}_n(ne)+\frac{(1-e^2)}{e^2}J^2_n(ne)\Big]\Big(1-\frac{n^2_0}{n^2}\Big)^\frac{1}{2}\Big(1+\frac{1}{2}\frac{n^2_0}{n^2}\Big),
\label{eq:k2}
\end{eqnarray}
and use these notations to rewrite Eq.~(\ref{eq:vecto}) as
\begin{equation}
\frac{dE}{dt}=\frac{g^2}{3\pi}a^2M^2\Big(\frac{Q_1}{m_1}-\frac{Q_2}{m_2}\Big)^2\Omega^4\Big(\frac{\Omega^2}{M^2_{Z^\prime}}K_1(n_0,e)+K_2(n_0,e)\Big).
\label{eq:vectory}
\end{equation}
This is the energy loss due to radiation of proca vector massive boson from NS-NS binaries. For NS-WD binaries, the energy loss is same as Eq.~(\ref{eq:vectory}) with $Q_2=0$ because white dwarfs do not have any muon charges. 
\section{Energy loss due to radiation of massive $L_\mu-L_\tau$  gauge boson}
If the $Z^\prime$ boson is a gauge field, then from gauge invariance $k_\mu J^\mu=0$ and, consequently, the second term in the polarization sum of Eq.~(\ref{eq:pol_sum}) will not contribute to the energy loss formula. Using the same procedure that has been described in the previous section, we obtain the rate of energy loss
\begin{equation}
\frac{dE}{dt}=\frac{g^2}{6\pi}a^2 M^2\Big(\frac{Q_1}{m_1}-\frac{Q_2}{m_2}\Big)^2 \Omega^4\sum_{n>n_0}2n^2\Big[J^{\prime^2}_n(ne)+\frac{(1-e^2)}{e^2}J^2_n(ne)\Big]\Big(1-\frac{n^2_0}{n^2}\Big)^\frac{1}{2}\Big(1+\frac{1}{2}\frac{n^2_0}{n^2}\Big).
\label{eq:vect}
\end{equation}
or
\begin{equation}
\frac{dE}{dt}=\frac{g^2}{3\pi}a^2 M^2\Big(\frac{Q_1}{m_1}-\frac{Q_2}{m_2}\Big)^2 \Omega^4 K_2(n_0,e),
\label{eq:avect}
\end{equation}
where $K_2(n_0,e)$ is defined earlier in Eq.~(\ref{eq:k2}). Since $K_2(n_0,e)_{n_0=0}\geq K_2(n_0,e)_{n_0\neq0}$ the massless limit gives the stronger bound on the energy loss.
This is the energy loss due to massive vector gauge boson radiation which has the similar form as previously obtained in \cite{krause}. Our method in obtaining the formula is different, where we can differentiate between the radiation rate of massive vector gauge bosons from the massive proca fields.

The rate of change of the orbital period due to energy loss is
\begin{equation}
\frac{dP_b}{dt}=-6\pi G^{-3/2}(m_1 m_2)^{-1}(m_1+m_2)^{-1/2}a^{5/2}\Big(\frac{dE}{dt}+\frac{dE_{GW}}{dt}\Big),
\label{eq:total}
\end{equation}
where $\frac{dE_{GW}}{dt}$ is the rate of energy loss due to quadrupole formula for the gravitational radiation and is given by \cite{peters}
\begin{equation}
\frac{dE_{GW}}{dt}=\frac{32}{5}G\Omega^6 M^2 a^4(1-e^2)^{-7/2}\Big(1+\frac{73}{24}e^2+\frac{37}{96}e^4\Big).
\label{eq:grav}
\end{equation}
In the massless limit of the vector gauge boson (i.e; $M_{Z^\prime}=0$ implies $n_0=0$), the rate of energy loss from Eq.~(\ref{eq:vect}) becomes
\begin{eqnarray}
\frac{dE}{dt}&=&\frac{g^2}{3\pi}a^2 M^2\Big(\frac{Q_1}{m_1}-\frac{Q_2}{m_2}\Big)^2 \Omega^4\sum^\infty_{n=1}n^2\Big[J^{\prime^2}_n(ne)+\frac{(1-e^2)}{e^2}J^2_n(ne)\Big]\nonumber\\
&=& \frac{g^2}{6\pi}a^2 M^2\Big(\frac{Q_1}{m_1}-\frac{Q_2}{m_2}\Big)^2 \Omega^4\frac{(1+\frac{e^2}{2})}{(1-e^2)^\frac{5}{2}}.
\end{eqnarray}
If the orbit is circular then the angular velocity is a constant over the orbital period and the Fourier expansion of the orbit contains only one term for $\omega=\Omega$. In an eccentric orbit the angular velocity is not constant and that means the Fourier expansion must sum over the harmonics $n\Omega$ of the fundamental.

In the following we will put constraints on the mass of the vector gauge boson and on the $L_\mu -L_\tau $ coupling constant from the decay of orbital period of four compact binary systems using Eq.~(\ref{eq:avect}), Eq.~(\ref{eq:total}) and Eq.~(\ref{eq:grav}).
\section{Constraints on gauge boson mass and its coupling for different compact binaries}
\subsection{PSR B1913+16: Hulse Taylor binary pulsar}

This was the first binary pulsar which was discovered by Hulse and Taylor in 1974 \cite{hulse,taylor,weisberg}. The observed value of the orbital period of PSR B1913+16 decays at the rate of $\dot{P_b}=2.4225\times 10^{-12}ss^{-1}$ and GR predicts its value as $\dot{P_b}=2.4025\times 10^{-12}ss^{-1}$ \cite{kt,krause}. The masses of the two neutron stars are $m_1=1.42  M_\odot$ and $m_2=1.4  M_\odot$. The orbit is highly eccentric $(e=0.617127)$ \cite{mohanty}. The average orbital frequency is $\Omega=1.48\times 10^{-28}GeV$. Massive gauge bosons can radiate from the neutron star if $n\Omega>M_{Z^\prime}$ where $n=1$ stands for the fundamental mode. This implies, for the radiation of gauge boson, the mass is constrained as $M_{Z^\prime}<1.48\times10^{-19}eV$. The semi-major axis $a$ of the orbit is obtained from Kepler's law $T^2=4\pi^2/G(m_1+m_2)a^3$ and it is $a=1.087\times 10^{25}GeV^{-1}$. Here $(Q_1/m_1-Q_2/m_2)= 10^{-4}GeV^{-1}$ where $Q=N$, $N$ is the number of muons which is roughly $10^{55}$ \cite{new}. From Eq.~(\ref{eq:avect}) it is clear that the radiation of vector gauge boson is possible if the charge to mass ratio is different for the two neutron stars. The contribution from the radiation of some vector gauge boson particles must be within the excess of the decay of the orbital period, i.e $\dot{P}_{b(vector)}\leqslant \vert \dot{P}_{b(observed)}- \dot{P}_{b(gw)}\vert$. Since $K_2(n_0,e)_{n_0=0}>K_2(n_0,e)_{n_0\neq0}$, the massless limit gives the stronger bound. We get the bound of the gauge boson coupling constant from the orbital decay period in the massless limit as 
\begin{equation}
g\leq 2.21\times 10^{-18}.
\end{equation}




From the fifth force constraint the ratio of the fifth force to the gravitational force should be less than unity which implies
\begin{equation}
\frac{g^2 N^2}{4\pi G m_1 m_2}\leq1.
\end{equation}
This gives the upper bound on $g$ as $g\leq4.99\times 10^{-17}$ for HT binary.
\subsection{PSR J0737-3039: Double binary pulsars}
The double binary pulsar system (PSR J0737-3039A/PSR J0737-3039B) \cite{kramer} consists of two neutron stars and both of them are pulsars emitting electromagnetic waves in the radio wavelength range. This compact binary system has an average orbital period $P_b=2.4h$. The masses of the two stars are $M_1=1.338M_\odot$ and $M_2=1.25M_\odot$, and the eccentricity of the orbit is $e=0.087$. Its observed orbital period decays at a rate $\dot{P_b}=1.252\times10^{-12}ss^{-1}$ whereas its expected value from GR is $1.24787\times10^{-12}ss^{-1}$. The orbital frequency is $\Omega=4.79\times 10^{-28}GeV$ and the semi-major axis of the orbit is $a=4.83\times10^{24}GeV^{-1}$. Here the difference between the charge to mass ratio is  $(Q_1/m_1-Q_2/m_2)=5.27\times 10^{-4}GeV^{-1}$. Since the massless limit gives the stronger bound, We get the bound of the muon gauge boson coupling constant in the massless limit as 
\begin{equation}
g\leq 2.17\times 10^{-19}. 
\end{equation}


From fifth force constraint we can write the upper bound on $g$ as $g\leq4.58\times 10^{-17}$ for PSR J0737-3039. 
\subsection{PSR J0348+0432: Pulsar white dwarf binary }
PSR J0348+0432 \cite{john} is a pulsar white dwarf binary system which consists of a pulsating NS and a low mass WD companion. The orbital period of this very low eccentric compact binary system is $P_b=2.46h$. The mass of the pulsar is $M_1=2.01M_\odot$ and the mass of the white dwarf is $M_2=0.172M_\odot$. The observed orbital period decay rate is $\dot{P_b}=0.273\times10^{-12}ss^{-1}$ and its GR predicted value is $\dot{P_b}=0.258\times10^{-12}ss^{-1}$. The semi major axis of the orbit is obtained from Kepler's law and it is $a=4.64\times 10^{24}GeV^{-1}$. The orbital frequency is $\Omega=4.67\times 10^{-19}eV$. Since the muon content in the white dwarf is negligible so $Q_2=0$ and $(Q_1/m_1-Q_2/m_2)=4.97\times 10^{-3}GeV^{-1}$. Since the massless limit gives the stronger bound in the mass scale $M_Z^\prime<10^{-19}$ eV, the muon gauge boson coupling in the massless limit is 
\begin{equation}
g\leq 9.02\times 10^{-20}.
\end{equation}

\subsection{PSR J1738+0333: Pulsar white dwarf binary }
PSR J1738+0333 \cite{paulo} is consist of a pulsar and a low mass white dwarf companion. The orbital period is $P_b=8.5h$. The eccentricity of the orbit is very small, $e<4\times 10^{-7}$. The mass of the pulsar is $M_1=1.46M_\odot$ and the mass of the white dwarf is $M_2=0.181M_\odot$. The intrinsic orbital period decay is $\dot{P_b}=-25.9\times10^{-15}ss^{-1}$ and its GR predicted value is $\dot{P_b}=-27.7\times10^{-15}ss^{-1}$. The semi major axis of the orbit is calculated from Kepler's law and it is $a=9.647\times 10^{24}GeV^{-1}$. The orbital frequency is $\Omega=1.35\times 10^{-19}eV$. Here $(Q_1/m_1-Q_2/m_2)=6.85\times 10^{-3}GeV^{-1}$. Since the massless limit gives the stronger bound in the mass scale $M_Z^\prime<10^{-19}$ eV, the muon gauge boson coupling in the massless limit is 
\begin{equation}
g\leq 4.24\times 10^{-20}.
\end{equation}

 
In TABLE \ref{tableI} we show the bounds on $g$ from fifth force and orbital period decay for the four compact binary systems in the mass range $M_Z^\prime<10^{-19}$ eV. Here we use the fact that the fifth force should be subdominant over gravity so as not to destabilize the Kepler's orbit. On the other hand, the bound on $g$ comes from the radiation of vector boson as described in detail in this paper. The constraint on the coupling from radiation loss is much more stringent compared to the fifth force constraint.

\begin{table}[h]
\caption{\label{tableI} Summary of the upper bounds on gauge boson-muon coupling $g$ for PSR B1913+16, PSR J0737-3039, PSR J0348+0432, and PSR J1738+0333. We take the mass regime as $M_Z^\prime<10^{-19}$ eV.}
\centering
\begin{tabular}{ lcc  }
 
 \hline
Compact binary system \hspace{0.5cm} & $g$(fifth force)\hspace{0.5cm} & $g$(orbital period decay)\\
 \hline
PSR B1913+16  & $\leq 4.99\times 10^{-17}$  & $\leq 2.21\times 10^{-18}$ \\
PSR J0737-3039 & $\leq 4.58\times 10^{-17}$  &$ \leq 2.17\times 10^{-19}$\\
PSR J0348+0432 & $ -$  & $\leq 9.02\times 10^{-20}$ \\
PSR J1738+0333 & $ -$  &$ \leq 4.24\times 10^{-20}$\\
 \hline
\end{tabular}
\end{table}
\begin{figure}[!htbp]
\centering
\subfigure[$g$ vs. $M_{Z^\prime}$ (gauge field)]{\includegraphics[width=4.0in,angle=360]{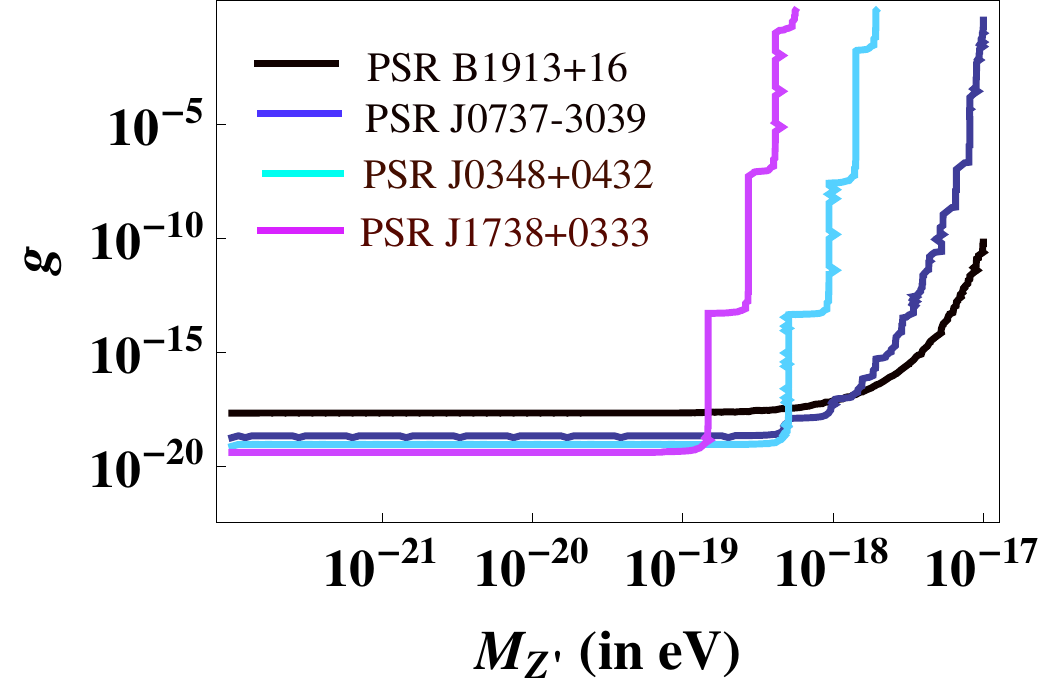}\label{subfig:vvsa3}}
\subfigure[$g$ vs. $M_{Z^\prime}$ (proca field)]{\includegraphics[width=4.0in,angle=360]{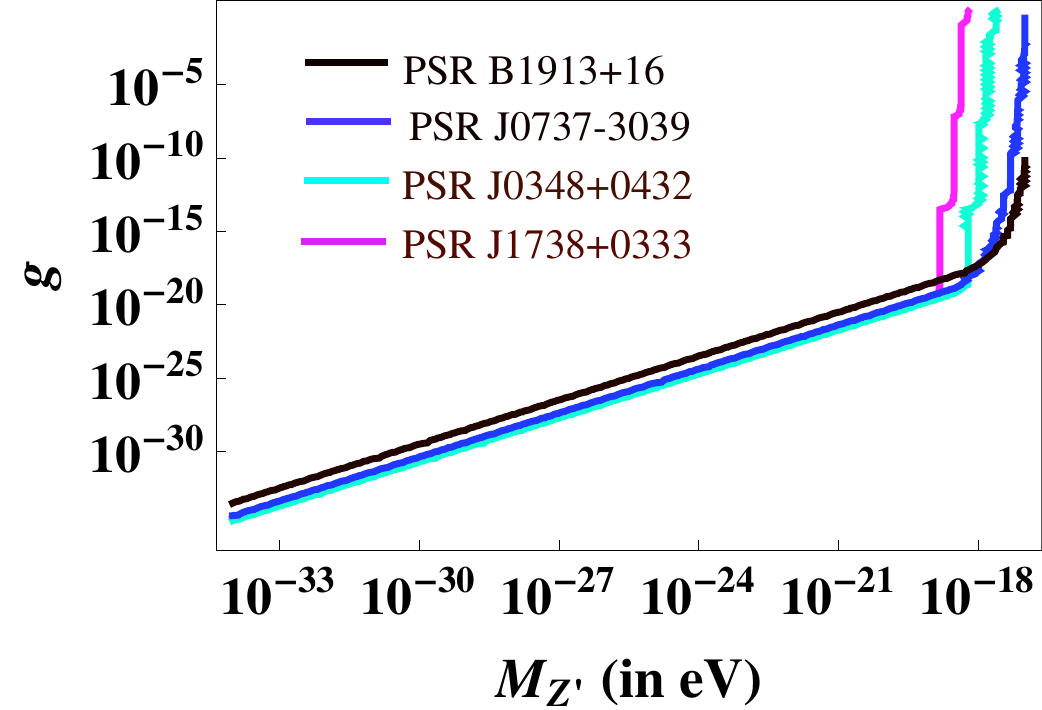}\label{subfig:vvsr4}}
\caption{(a) Exclusion plots to constrain the coupling of the gauge field and (b) the proca field in a gauged $L_\mu-L_\tau$ scenario for four compact binary systems. The regions above the coloured lines are excluded.}
\label{fig:PSRJ}
\end{figure}
In FIG.\ref{fig:PSRJ} we show the exclusion plots to constrain the coupling $g$ for the gauge field and the proca field using Eq.~(\ref{eq:vectory}) in a gauged $L_\mu-L_\tau$ scenario for four compact binary systems. The regions above the coloured lines are excluded for the corresponding binary systems. Here, larger parameter space of $g$ is excluded for the proca field. There is a $1/M^2_{Z^\prime}$ term in the polarization sum of massive vector bosons (both proca and gauge bosons). Due to gauge invariance this term does not contribute in the gauge boson calculation but it is present in the proca field calculation and we compare the radiation of the proca and the gauge fields separately in the plots. The regions above the coloured lines are excluded for the corresponding binary systems.

FIG.\ref{subfig:vvsa3} shows for gauge boson the coupling $g$ is almost constant in the mass range $M_Z^\prime<10^{-19}$eV. The coupling $g$ will increase with $M_{Z^\prime}$ in the mass range $M_{Z^\prime}>10^{-19}eV$, as only higher modes $(n>n_0>1)$ contribute to $K_2(n_0,e)$. For low eccentric binary orbits, the rise in $g$ with respect to $M_{Z^\prime}$ is sharp. Note that  for circular binary orbit only the $n=1$ mode can contribute. As a result for $M_{Z^\prime}>\Omega$, there is no constraint on $g$. 

In FIG.\ref{subfig:vvsr4}, $g$ varies linearly with respect to $M_{Z^\prime}(<10^{-19}$eV) due to the contribution of $\Omega^2/M^2_{Z^\prime}K_1(n_0,e)$ term for the proca field. We obtain the upper bounds on $M_{Z^\prime}/g$ for a proca field in the small $M_{Z^\prime}$ limit. From the orbital period decay, we get for PSR B1913+16, $M_{Z^\prime}/g\leq 0.306eV$, for PSR J0737-3039, it is $M_{Z^\prime}/g\leq 2.307eV$, for PSR J0348+0432, the bound is $M_{Z^\prime}/g\leq 5.13eV$ and, for PSR J1738+0333, the bound is $M_{Z^\prime}/g\leq 3.19eV$. 
\section{Discussions}
Due to the presence of significant number of muons in the neutron stars, we can put bounds on the ultra light vector gauge boson mass in the $L_\mu-L_\tau$ gauge and on the gauge coupling from the observations of orbital period decay of the four compact binary systems. Mainly the gravitational quadrupole radiation contributes to the decay in orbital period. The radiation by other ultra light particles also contribute to the orbital period decay to less than $1\%$. From the decay of orbital period, we obtain the $L_\mu-L_\tau$ gauge coupling for PSR B1913+16 as $g\leq 2.21\times 10^{-18}$, for PSR J0737-3039, it is $g\leq 2.17\times 10^{-19}$, for PSR J0348+0432, the coupling is $g<9.02\times10^{-20}$ and, for PSR J1738+0333, the coupling is $g<4.24\times10^{-20}$ in the massless limit and is true up to $M_Z^\prime<10^{-19}$ eV. Due to the fact of $K_2(n_0,e)_{n_0=0}\geq K_2(n_0,e)_{n_0\neq0}$, the massless limit gives the stronger bound for the radiation of massive vector gauge boson. The radiation of vector gauge boson particles is possible if the charge to mass ratio is different for two neutron stars. We have shown the exclusion plots of $g$ vs $M_{Z^\prime}$ for the radiation of massive vector gauge boson and proca field from the NS-NS and NS-WD binaries. The main uncertainty of the gauge coupling bound comes from the number of muons in the neutron star which depends on different QCD equation of states.


\begin{thebibliography}{100} 
\bibitem{hulse}R.A. Hulse and J.H. Taylor, Astrophys.J.Lett \textbf{195}, L51 (1975).
\bibitem{taylor}J.H. Taylor and J.M. Weisberg, Astrophys.J.\textbf{253}, 908 (1982).
\bibitem{weisberg}J.M. Weisberg and J.H. Taylor, Phys. Rev. Lett.\textbf{52},1348 (1984).
\bibitem{peters}P.C. Peters and J. Mathews, Phys. Rev. \textbf{131}, 435 (1963).
\bibitem{mohanty}S. Mohanty and P.K. Panda, Phys.Rev.D \textbf{53},5723.
\bibitem{hook}A. Hook and J. Huang, JHEP06(2018)036.
\bibitem{mairi}J. Huang, M. C. Johnson, L. Sagunski, M. Sakellariadou, J. Zhang, Phys. Rev. D {\bf 99}, 063013 (2019).
\bibitem{tanmay}T.K. Poddar, S. Jana and S. Mohanty, arXiv: 1906.00666.
\bibitem{newre}J.M. Weisberg and J.H. Taylor, ASP Conf.Ser. 328 (2005) 25.
\bibitem{mahapatra} I. Goldman, R. N. Mahapatra, S. Nussinov, arXiv: 1901.07077.
\bibitem{foot}R. Foot, Mod. Phys. Lett. \textbf{A}\textbf{6}, 527(1991)
\bibitem{he}Xiao-Gang He, G. C. Joshi, H. Lew, and R. R. Volkas, Phys. Rev D {\bf 44}, 2118 (1991).
\bibitem{r} R. Foot, X.-G. He, H. Lew, and R. R. Volkas, Phys. Rev. D {50}, 4571 (1994).
\bibitem{heeck}J. Heeck and W. Rodejohann, Phys. Rev. \textbf{D84} (2011) 075007.
\bibitem{masso}J. A. Grifols, E. Masso, Phys.Lett. B579 (2004) 123-126.
\bibitem{anjan}A. S. Joshipura, S. Mohanty, Phys. Lett. B584 (2004) 103-108.
\bibitem{amol} A. Bandyopadhyay, A. Dighe, A. S. Joshipura, Phys. Rev. D {\bf 75}, 093005 (2007).
\bibitem{agarwalla}M. Bustamante, S.K.Agarwalla, Phys. Rev. Lett. {\bf 122}, 061103 (2019).
\bibitem{dutt}J. M. Pearson, N. Chamel, A. Y. Potekhin, A. F.
Fantina, C. Ducoin, A. K. Dutta, and S. Goriely,
“Unified equations of state for cold non-accreting
neutron stars with BrusselsMontreal functionals  I. Role
of symmetry energy,”
Mon. Not. Roy. Astron. Soc.
481
no. 3, (2018) 2994–3026, [
1903.04981
].
\bibitem{potekhin}A. Y. Potekhin, A. F. Fantina, N. Chamel, J. M. Pearson and S. Goriely, Astron, Astrophys. \textbf{560}(2013)A48.
\bibitem{goriely}S. Goriely, N. Chamel and J. M. Pearson, Phys. Rev. C \textbf{88}, 024308(2013) .
\bibitem{chamel}S. Goriely, N. Chamel and J. M. Pearson, Phys. Rev. C \textbf{82}, 035804 (2010).
\bibitem{pearson}J. M. Pearson, N. Chamel, A. Y. Potekhin, A. F. Fantina, C. Ducoin, A. K. Dutta, S. Goriely, MNRAS 481, 2994-3026 (2018); erratum: MNRAS 486, 768 (2019).
\bibitem{new}R. Garani, J. Heeck,  	arXiv:1906.10145.
\bibitem{garani}R. Garani, Y. Genolini and T. Hambye, JCAP \textbf{1905}(2019)035.
\bibitem{bell}N.F. Bell, G. Busoni and S. Robes ,arXiv:1904.09803.
\bibitem{kramer} M. Kramer et al, Science \textbf{314},97 (2006).
\bibitem{john} John Antoniadis et al., Science  26 Apr 2013: Vol. 340, Issue 6131, 1233232.
\bibitem{paulo} Paulo C.C Freire et al., The relativistic pulsar-white dwarf binary PSR J1738+0333-II. The
most stringent test of scalar-tensor gravity.,Mon. Not. R. Astron. Soc. 423, 33283343 (2012).
\bibitem{feng}F. S. Zhang, L. W. Chen, Chinese Physics Letters 18 (1), 142-144 (2001).
\bibitem{krause}D. E. Krause, H. T. Kloor, and E. Fischbach, Phys. Rev. D 49, 6892.
\bibitem{kt}J.H.Taylor,Class.Quantum Grav.10, S167-S174 (1993).
\end{thebibliography}
\end{document}